\documentclass[twocolumn]{pasj00}

\begin{document}
\SetRunningHead{Lee et al.}{Incidence of high-amplitude $\delta$ Scuti stars}
\Received{}
\Accepted{}

\title{Incidence of high-amplitude $\delta$ Scuti type variable stars}
\author{Yong-Hwa \textsc{Lee},
Sungsoo S. \textsc{Kim},\thanks{Corresponding author; sungsoo.kim@khu.ac.kr}
Jihye \textsc{Shin},
Jiwon \textsc{Lee}}
\affil{Department of Astronomy \& Space Science, Kyung Hee University,
Yongin, Kyunggi 446-701, Korea}
\and
\author{Ho Jin}
\affil{Korea Astronomy \& Space Science Institute, 61-1 Hwa-Am, Yuseong,
Daejeon 305-348, Korea}


\maketitle

\begin{abstract}
An order-of-magnitude estimate for the incidence of high-amplitude $\delta$
Scuti-type variable stars (HADS) in the $\delta$ Scuti area of the H-R
diagram is calculated.
Using a model for the stellar distribution in the Milky Way, we calculate the
number of stars that are expected to fall in the $\delta$ Scuti area of the
H-R diagram within the magnitude range and sky coverage of the ROTSE Survey
for Variables I (RSV1).  The incidence of the HADS phenomenon
is then obtained by comparing the number of stars calculated by the model
and the actual, observed number of HADS in the RSV1.
We find that $\sim 0.3$~\% of the stars that lie in the $\delta$ Scuti area of
the H-R diagram within the RSV1 observational limits exhibit the HADS
phenomenon.  This number is much lower than the incidence of the low-amplitude
$\delta$ Scuti stars (LADS), $\gtrsim 1/3$, implying that the HADS phenomenon
takes place in a very small fraction of stars and/or its duration is very
short, compared to the LADS.
\end{abstract}

\section{Introduction}
\label{sec:introduction}

$\delta$ Scuti stars are pulsating variables located in the lower part of the
classical Cepheid instability strip on and just above the main sequence in
the Hertzsprung-Russell (H-R) diagram with spectral types roughly from
A2 to F2 and periods between $\sim 0.04$ and $\sim 0.25$ days.
They are astrophysically interesting variables as their pulsations may
give important information about their asteroseismological properties.
While the low-amplitude $\delta$ Scuti stars (LADS) tend to be nonradial
pulsators with high rotational velocities, the high-amplitude $\delta$ Scuti
stars (HADS) are mostly radial pulsators with small rotational velocities
(Breger 2000, McNamara 2000a).  Following Breger (2000), we adopt
$\Delta m_V > 0.3$ mag for the definition of the HADS.

An immense amount of scientific researches concerning $\delta$ Scuti stars have
taken place during the past few decades (see the references in Breger 2000).
Rodriguez \& Breger (2001) showed statistical results of the physical
properties of 636 $\delta$ Scuti stars discovered by individual and all-sky
survey(s), but HADS are still relatively less studied variables.
For example, between 1/3 and 1/2 of the stars located in the lower
instability strip are known to show phetometrically detectable light
variability due to pulsation with an amplitude limit between 0.003 and
0.010 mag (Breger 2000), and $\sim 1/4$ of the solar-neighborhood stars
in the lower instability strip is found to be variable (Poretti et al. 2003),
but the incidence of HADS phonomenon is almost unknown.  HADS are
particularly interesting variables as they show relatively large magnitude
variations compared to their short periods.  Information on the incidence
of these variables would help us better understand the nature and evolution
of this type of pulsating stars.

In the present paper, we calculate the number of stars that are expected to
fall in the $\delta$ Scuti area of the H-R diagram for
the magnitude range and sky coverage of the ROTSE (Robotic Optical Transient
Search Experiment) Survey for Variables I (RSV1; Akerlof et al. 2000) catalog.
By comparing this number to the actual, observed number of HADS
in the RSV1, we estimate the occurrence frequency of HADS in the
$\delta$ Scuti area of the H-R diagram.

\section{The ROTSE Survey for Variables I}
\label{sec:RSV1}

Observations used for the present study were taken by ROTSE-I, an array of
four telescopes each with an $8\fdg2$ $\times$ $8\fdg2$ field of view
(Akerlof et al. 2000).  The ROTSE telescope was operated without any filters
so the spectral response is primarily limited by the sensitivity of the CCD.
Designed to find optical counterparts to gamma-ray bursts, it spent most of its
time automatically surveying the entire visible sky twice each night.

While no filters were used for the ROTSE observations, the star count model
adopted in the present work involves magnitudes in conventional photometric
systems.  Thus transformation from a certain photometric system to the
ROTSE magnitude is necessary.  We adopt the empirical transformation from
the $B$ and $V$ magnitudes to the ROTSE magnitude that Akerlof et al. (2000)
obtained by comparing the Tycho catalog (ESA 1997) and ROTSE observations:
\begin{equation}
	m_{ROTSE} = 0.5 + m_V - \frac{m_B - m_V}{1.875}
\end{equation}
A constant of 0.5 was inserted here to compensate the magnitude
discrepancy between the Tycho and ROTSE data (P. R. Wo\'zniak 2005, private
communication).

The RSV1 is the result of analyzing 5~\% of the available ROTSE all-sky data
over a period of 3 months (March 15 to June 15, 1999).  It covers a Galactic
latitude range of $b=\timeform{13D}-\timeform{88D}$ and a longitude range of
$l=\timeform{10D}-\timeform{80D}$.  Out of the 1781 variable stars detected
(mean magnitudes between $m_{ROTSE}=10.0$ and 15.5~mag), 91 variables,
of which three are known from the General Catalog of Variable Stars
(GCVS; Kholopov 1998), have been classified into the $\delta$ Scuti type
by the RSV1.

However, some of the $\delta$ Scuti stars have light curves (or, periods and
amplitudes) similar to those of W Ursae Majoris (W UMa) type eclipsing
binaries, which are often misclassified into $\delta$ Scuti type pulsating stars
particularly when the number of observing epochs are not enough or
less rigorous criteria are used in automated classifications.
Poretti (2001) and Morgan (2003) showed that Fourier decomposition can
be an effective way of identifying some pulsating stars such as HADS,
RRc, and RRab stars.

Jin et al. (2003, 2004) conducted follow-up observations for 49 $\delta$
Scuti stars in the RSV1 catalog whose Fourier parameters show $\delta$ Scuti
signatures, and found that only 6 of them (see Table \ref{table:bonafide})
are bona fide $\delta$ Scuti stars and most of the rest are W UMa stars (note
that all of the 6 bona fide $\delta$ Scuti stars are HADS).  Judging from the
Fourier parameters of the remaining 42 stars, it is quite unlikely that
there are additional bona fide HADS in the RSV1 catalog.  Thus we
conclude that the number of HADS in the RSV1 survey is $\sim 6$, and this
number will be compared in \S \ref{sec:comparison} to the number of $\delta$
Scuti candidates in the magnitude and spatial coverage of the RSV1 in order
to determine the incidence of the $\delta$ Scuti phenomenon.
We choose the RSV1 survey for this comparison as its $\delta$ Scuti content
has been thoroughly studied and the number of bona fide HADS in the survey
is reliably given.

\begin{table*}
\caption{Bona Fide $\delta$ Scuti Stars from RSV1
\label{table:bonafide}}
\centering
\begin{tabular}{c c c c c}
\hline\hline
         & Galactic Latitude & $m_{ROTSE}$ & Period & Amplitude \\
ROTSE ID & (degree)          & (mag)       & (day)  & (mag)     \\
\hline
J152406.95+365200.9 & 56.4 & 10.99 & 0.104 & 0.429 \\
J163117.94+115952.4 & 36.5 & 10.86 & 0.149 & 0.437 \\
J164839.21+302745.6 & 38.5 & 13.49 & 0.131 & 0.409 \\
J182943.22+280955.2 & 16.8 & 12.70 & 0.146 & 0.393 \\
J183206.54+403555.9 & 20.7 & 12.87 & 0.102 & 0.625 \\
J193445.28+455416.9 & 12.2 & 11.81 & 0.177 & 0.402 \\
\hline
\end{tabular}
\begin{list}{}{}
\item Note---From Jin et al. (2003).
\end{list}
\end{table*}

\section{Observational Completeness}
\label{sec:complete}

As the discovery of variable stars is not the primary goal of the
ROTSE-I experiment, the RSV1 catalog may not be complete.  Indeed,
the number of available epochs during the three-month period for each star
ranges from only $\sim 40$ to $\sim 110$.  However, Akerlof et al.
(2000) found that more than 80~\% of the RR Lyrae and $\delta$ Scuti stars
of the GCVS in the RSV1 survey area with $m_V$ between 10 and 13~mag are
recovered in the RSV1 catalog.

We have conducted a similar test with the All Sky Automated Survey (ASAS;
Pojmanski 2002) data that overlap the RSV1 catalog in magnitude range and
sky coverage.  The latest catalog of the ASAS (ASAS-3 Catalog of Variables
Stars) contains almost 8,000 periodic pulsators and over 10,000 eclipsing
binaries located in the southern hemisphere below the declination
$+28^\circ$ with $m_V < 14$~mag.  Inspections between RSV1 variables and
ASAS $\delta$ Scuti stars with $\Delta m_V>0.3$~mag and periods less than
0.3~days led to 20 ASAS $\delta$ Scuti stars in the RSV1 area.  Out of the 20,
11 stars are detected as variables by the RSV1, resulting in a recovery
fraction of 55~\%.

We conclude that the detection efficiency of the HADS in the RSV1 survey is
at least $\sim 50$~\% for $m_V<14$~mag.


\section{The Star Count Model of the Milky Way}
\label{sec:model}

A detailed model by Wainscoat et al. (1992) for the point source sky that
comprises geometrically and physically realistic representations of the
galactic disk, bulge, stellar halo, spiral arms, molecular ring and the
extragalactic sky is used to calculate the number of stars that fall in
the $\delta$ Scuti area of the H-R diagram, i.e., stars with a luminosity
class of V and a spectral type between A6 and F0.\footnote{While the
effective temperatures of both LADS and HADS range from $\sim 7000$~K to
$\sim 8500$~K, which roughly corresponds to a spectral range of A2 through
F2, HADS have a narrower temperature range, $\sim 7200$~K through
$\sim 8000$~K (McNamara 2000b), which roughly corresponds to a spectral
range of A6 through F0.}  In this model, galactic
components are represented by 87 types of the Galactic source,
each fully characterized by scale heights, space densities, and absolute
magnitudes at $BVJHK$, $12 \mu$, and $25 \mu$.

Out of the 87 source types considered in the star count model, two
correspond to the HADS: A2--5~V, and F0--5~V.  Considering
the number of stars within 100~pc from the Sun as a function of spectral
type, we apply weights of 0.16 and 0.088 respectively to the forementioned
source types.  We use the numbers of stars within 100~pc from the Sun for
determining the weights because the adopted star count model gives the number
density of each source type as a function of distance and direction
from the Sun relative to the number density in the solar neighborhood.

The weights of 0.16 and 0.088 are obtained as follows.  The source types
neighboring to A2--5~V and F0--5~V in the Wainscoat et al. model are
B8--A0~V and F8~V.  Thus we assume that the A2--5~V type actually covers
spectral types A2~V through A7~V and a half of A1~V while the F0--5~V type
covers A8~V through F6~V.  Table~\ref{table:hipparcos} shows the numbers
of stars with spectral types between A1V and F6V within 100~pc from the Sun
in the Hipparcos catalog (ESA 1997).  We find from this Table that the
fraction of A6 through A7 to A1 through A7 (with a half weight for A1) is
0.16 while that of A8 though F0 to A8 through F6 is 0.088.

\begin{table*}
\caption{Numbers of Stars within 100~pc from the Sun
\label{table:hipparcos}}
\centering
\begin{tabular}{c c c c c}
\hline\hline
Spectral Type & Number of Stars & & Spectral Type & Number of Stars \\
\hline
A1~V &  91 & & A9V &  51 \\
A2~V &  98 & & F0V & 167  \\
A3~V & 116 & & F1V &  16  \\
A4~V &  40 & & F2V & 179  \\
A5~V &  55 & & F3V & 329  \\
A6~V &  13 & & F4V &  55  \\
A7~V &  55 & & F5V & 594  \\
A8~V &  23 & & F6V & 477  \\
\hline
\end{tabular}
\begin{list}{}{}
\item Note---From the Hipparcos catalog (ESA 1997).
\end{list}
\end{table*}

\section{Model-Data Comparison}
\label{sec:comparison}

The applicability of the adopted star count model to the RSV1 data
was tested by comparing the stellar luminosity functions (LFs) predicted
by the model to the observed LFs from the ROTSE survey, towards several
positions within the spatial coverage of the RSV1.  For this comparison,
we use the Nothern Sky Variability Survey (NSVS; Wo\'zniak et al. 2004)
data set, which is a collection of all (including the non-variable objects)
photometric data from the ROTSE-I survey.

\begin{figure*}
\centering
\FigureFile(14cm,14cm){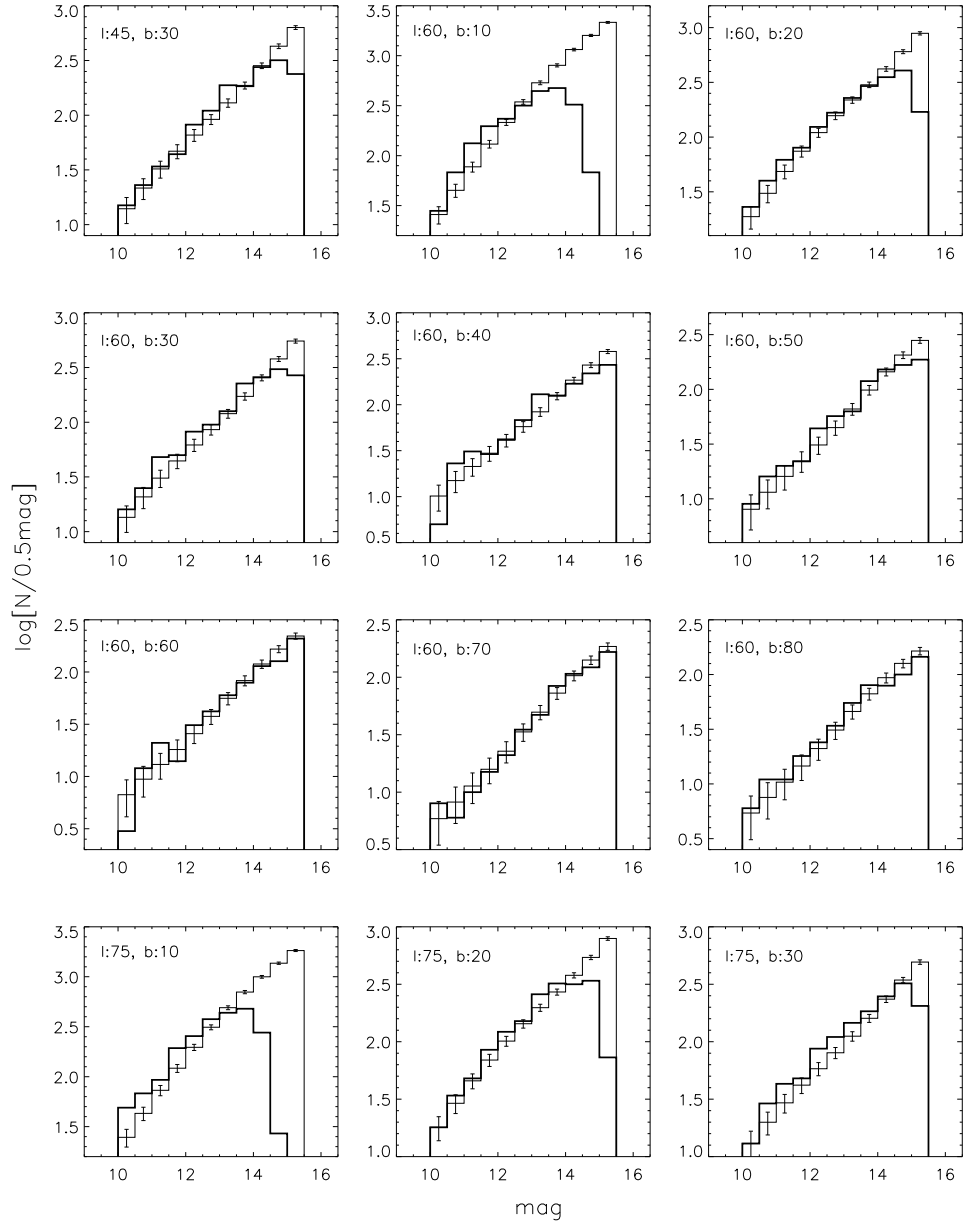}
\caption{\label{fig:hist1}Luminosity functions in ROTSE magnitude from
the model ({\it thin lines}) and the NSVS data ({\it thick lines}) towards
12 circular regions of a \timeform{1D} radius within the spatial coverage of
the RSV1.  The histograms are shown for the magnitude range of the NSVS,
$10 \leq m_{ROTSE} \leq 15.5$.}
\end{figure*}

Figure \ref{fig:hist1} compares the LFs from the model and the NSVS data
towards 12 positions within the spatial coverage of the RSV1.  The LFs are
shown for the magnitude range of the NSVS, $10 \leq m_{ROTSE} \leq 15.5$.
The incompleteness of the NSVS generally increases sharply near 15 mag,
but it starts at brighter magnitudes at $|b| < 20^\circ$ due to significant
stellar blending (Wo\'zniak et al. 2004).  These phenomena are well seen in
Figure \ref{fig:hist1}, and we limit our calculation to a magnitude range
of $10 \leq m_{ROTSE} \leq 14$ and a galactic latitude range of
$|b| > 20^\circ$.  Other than the discrepancy due to the incompleteness,
Figure \ref{fig:hist1} shows a generally good agreement between the model
and the data at $|b| > 20^\circ$, demonstrating that the adopted star count
model can be reliably used for estimating the number of stars in the
$\delta$ Scuti region of the H-R diagram.

We find that for the magnitude range of $10 \leq m_{ROTSE} \leq 14$ and
the spatial coverage of the RSV1 with a constraint $|b| > 20^\circ$,
the star count model yields a total of 280,210 stars, while it yields
2,441 HADS candidate stars, i.e., stars with spectral types
from $A6V$ to $F0V$.  As discussed in \S \ref{sec:RSV1}, there appear
to be 6 bona fide HADS in the RSV1 catalog, but only 4 of them satisfy
$|b| > 20^\circ$ and $m_{ROTSE} \le 14$~mag (see Table \ref{table:bonafide}).
Since the detection
efficiency of the HADS in the RSV1 survey is estimated to be $\sim 50$~\%
(see \S \ref{sec:complete}), the incidence of the HADS pheonomenon
in the $\delta$ Scuti region of the H-R diagram becomes $4/0.5/2,441 \simeq
0.33$~\%.  Considering only the Poisson noises, the relative uncertainty
involved with our estimate for the incidence is $\sim 50$~\%.  This
uncertainty is rather large, but our calculation certainly gives an
order-of-magnitude estimate for the incidence.

Our estimate for the incidence of the HADS, 0.3~\%, is much lower than that
of the LADS, $\sim 1/3$, implying that the HADS phenomenon takes place in
a very small fraction of stars and/or its duration is very short, compared
to the LADS.  As the relations between HADS and LADS in terms of their
origins and natures are still uncertain, our estimate for the incidence
of HADS in the $\delta$ Scuti region of the H-R diagram will be useful in
revealing their relations to the LADS.

We are grateful to M. Breger, C. Akerlof, P. R. W\'ozniak, Seung-Lee Kim,
G. Pojmanski, and B. Paczynski for their valuable discussions and help.
We thank the anonymous referee for his/her helpful comments, which greatly
improved our manuscript.
This publication makes use of the data from the Northern Sky Variability
Survey created jointly by the Los Alamos National Laboratory and University
of Michigan. The NSVS was funded by the Department of Energy, the National
Aeronautics and Space Administration, and the National Science Foundation.
This work was supported by the research fund from Kyung Hee University.


\end{document}